\documentclass[sigconf,screen,british]{acmart}

\AtBeginDocument{%
  \providecommand\BibTeX{{%
    Bib\TeX}}}

\usepackage[utf8]{inputenc} 
\usepackage[T1]{fontenc} 
\usepackage[british]{babel}

\usepackage[capitalise]{cleveref}
\Crefname{section}{Section}{Sections}
\crefname{section}{Sec.}{Secs.}
\Crefname{figure}{Figure}{Figures}
\crefname{figure}{Figure}{Figures}
\Crefname{table}{Table}{Tables}
\crefname{table}{Table}{Tabs.}

\usepackage{pifont}
\newcommand{\cmark}{\ding{51}}

\usepackage{xcolor}

\usepackage{esint} 
\usepackage{commath} 
\usepackage{units} 

\usepackage[skip=0pt]{caption} 
\usepackage{titlesec}
\titlespacing*{\section}{0pt}{0.1\baselineskip}{0.2\baselineskip}
\titlespacing*{\subsection}{0pt}{0.1\baselineskip}{0.2\baselineskip}
\titlespacing*{\subsubsection}{0pt}{0.1\baselineskip}{0.2\baselineskip}
\usepackage{comment}
\excludecomment{commented}

\usepackage{subcaption}
\usepackage{booktabs}
\usepackage{array} 
\usepackage{rotating}
\usepackage{multirow}
\usepackage{multicol}

\usepackage[inline]{enumitem} 
\setlist{nolistsep} 
\usepackage{todonotes}
\setcopyright{rightsretained}
\copyrightyear{2023}
\acmYear{2023}
\acmDOI{10.1145/3577190.3616120}

\acmConference[ICMI '23]{INTERNATIONAL CONFERENCE ON MULTIMODAL INTERACTION}{October 9--13, 2023}{Paris, France}

\acmBooktitle{INTERNATIONAL CONFERENCE ON MULTIMODAL INTERACTION (ICMI '23), October 9--13, 2023, Paris, France}
\acmISBN{979-8-4007-0055-2/23/10}

\makeatletter 
\tikzstyle{inlinenotestyle} = [
    draw=\@todonotes@currentbordercolor,
    fill=\@todonotes@currentbackgroundcolor,
    line width=0.5pt,
    inner sep = 0.8 ex,
    rounded corners=4pt]

\renewcommand{\@todonotes@drawInlineNote}{%
        {\begin{tikzpicture}[remember picture,baseline=(current bounding box.base)]%
            \draw node[inlinenotestyle,font=\@todonotes@sizecommand, anchor=base,baseline]{%
              \if@todonotes@authorgiven%
                {\noindent \@todonotes@sizecommand \@todonotes@author:\,\@todonotes@text}%
              \else%
                {\noindent \@todonotes@sizecommand \@todonotes@text}%
              \fi};%
           \end{tikzpicture}}}%
\makeatother

\begin{document}

\title{The {GENEA} {C}hallenge 2023: {A} large-scale
evaluation of gesture generation models in
monadic and dyadic settings}


\author{Taras Kucherenko}
\authornote{Equal contribution and joint first authors.}
\email{tkucherenko@ea.com}
\affiliation{
  \institution{SEED -- Electronic Arts (EA), Sweden}
  \city{Stockholm}
  \country{Sweden}
  }

\author{Rajmund Nagy}
\authornotemark[1]
\email{rajmund@kth.se}
\affiliation{%
  \institution{KTH Royal Institute of Technology}
  \city{Stockholm}
  \country{Sweden}
}
  
\author{Youngwoo Yoon}
\authornotemark[1]
\email{youngwoo@etri.re.kr}
\affiliation{%
  \institution{ETRI}
  \city{Daejeon}
  \country{Republic of Korea}
}
 
\author{Jieyeon Woo}
\email{woo@isir.upmc.fr}
\affiliation{%
 \institution{ISIR -- Sorbonne University}
 \city{Paris}
 \country{France}
 }
 
\author{Teodor Nikolov}
\email{tnikolov@hotmail.com}
\affiliation{%
 \institution{Umeå University}
 \city{Umeå}
 \country{Sweden}
 }

\author{Mihail Tsakov}
\email{tsakovm@gmail.com}
\affiliation{%
 \institution{Umeå University}
 \city{Umeå}
 \country{Sweden}
 }

\author{Gustav Eje Henter}
\email{ghe@kth.se}
\affiliation{%
  \institution{KTH Royal Institute of Technology}
  \city{Stockholm}
  \country{Sweden}
}
\newcommand{\toremove}[1]{\textbf{\textcolor{red}{#1}}}
\newcommand{\mayberemove}[1]{\textcolor{brown}{\textbf{#1}}}
\newcommand{\rn}[1]{\textcolor{orange}{#1}}
\newcommand{\moved}[1]{\textbf{\textcolor{yellow}{#1}}}
\renewcommand{\shortauthors}{Kucherenko, Nagy, Yoon, et al.}

\begin{abstract}
This paper reports on the GENEA Challenge 2023, in which participating teams built speech-driven gesture-generation systems using the same speech and motion dataset, followed by a joint evaluation. 
This year's challenge provided data on both sides of a dyadic interaction, allowing teams to generate full-body motion for an agent given its speech (text and audio) and the speech and motion of the interlocutor. 
We evaluated 12 submissions and 2 baselines together with held-out motion-capture data in several large-scale user studies. The studies focused on three aspects: 1) the human-likeness of the motion, 2) the appropriateness of the motion for the agent's own speech whilst controlling for the human-likeness of the motion, and 3) the appropriateness of the motion for the behaviour of the interlocutor in the interaction
, using a setup that controls for both the human-likeness of the motion \emph{and} the agent's own speech. 
We found a large span in human-likeness between challenge submissions, with a few systems rated close to human mocap. Appropriateness seems far from being solved, with most submissions performing in a narrow range slightly above chance, far behind natural motion. The effect of the interlocutor is even more subtle, with submitted systems at best performing barely above chance.
Interestingly, a dyadic system being highly appropriate for agent speech does not necessarily imply high appropriateness for the interlocutor. 
Additional material is available via the project website at \href{https://svito-zar.github.io/GENEAchallenge2023/}{svito-zar.github.io/GENEAchallenge2023/}.
\end{abstract}




\maketitle

\section{Introduction}

Verbal communication is not the only way to convey messages; non-verbal behaviour is also a significant factor in human communication \cite{mcneill1992hand}. Much of non-verbal behaviour consists of hand gestures and body gestures, which closely relate to speech content and have been shown to improve speech understanding \cite{holler2018processing}. 

Automatic gesture generation for embodied conversational agents (ECAs) has been studied for more than two decades. 
Synthetic gestures were initially created using rule-based systems, e.g., \cite{cassell2001beat,salvi2009synface}, but data-driven approaches that learn from human motion data 
have recently emerged \cite{bergmann2009GNetIc,levine2010gesture,chiu2015predicting,yoon2019robots,kucherenko2020gesticulator,alexanderson2020style,yoon2020speech}. 
For more in-depth reviews of speech gesture generation, refer to \citet{nyatsanga2023comprehensive}.

The present study focuses on a fair, systematic comparison of
systems that automatically generate non-verbal behaviour.
By comparing different methods and evaluating their effectiveness, it is possible to accurately assess and improve the current state of the art. Additionally, this comparison helps to identify important aspects of gesture generation and where the main issues lie. We accomplished this by running an open challenge, the GENEA\footnote{GENEA stands for ``Generation and Evaluation of Non-verbal Behaviour for Embodied Agents''.} Challenge 2023. 

In this paper, we report on the challenge, in which we provided the same dataset, evaluation criteria, and visualisation process to ensure that all major sources of variation are accounted for, except for a gesture-generation system itself. 
The GENEA Challenge 2023 builds on previous GENEA Challenges \cite{kucherenko2021large,yoon2022genea}, especially the 2022 challenge described in \citet{yoon2022genea} and \citet{kucherenko2023evaluating}.
In addition to being an exercise in benchmarking both new and previously-published gesture-generation methods, the results of the two previous challenges have since helped improve gesture-generation benchmarking in other ways as well. Researchers have, for example, used the data splits \cite{kim2023mpe4g, alexanderson2023listen}, the visualisation \cite{wang2021integrated}, and the objective \cite{bhattacharya2021speech2affectivegestures} and subjective \cite{yoon2021sgtoolkit, yang2023diffusestylegesture, yang2023qpgesture} evaluation methodologies, as a basis for future research. The data has also been used to benchmark subsequent gesture-generation models \cite{ferstl2021expressgesture, yazdian2022gesture2vec}, and even for automatic quality assessment \cite{he2022automatic}. 
The main difference from previous GENEA Challenges is that we considered gesture generation for dyadic scenarios (meaning considering both conversation partners) and evaluated the interlocutor appropriateness of submitted systems.

Our contributions include: 1) providing common elements of dataset, visualisation, and evaluation setup for a fair comparison between gesture generation systems; 2) organising an open challenge to promote the development of the field and encourage researchers to share their ideas; and 3) three large-scale user studies for human-likeness of motion, speech appropriateness, and interlocutor appropriateness that evaluate 14 gesture-generation systems (including two baselines).

\begin{commented}
\section{Related Work}
\mayberemove{Human gesture perception is highly subjective, and there are currently no widely accepted objective measures of gesture perception,
so the golden standard of evaluation in the gesture-generation field is a human assessment \cite{wolfert2021review}.
Unfortunately, particular model families are only applied to particular datasets, and never contrasted against one another.
}
\mayberemove{Other fields have done well using challenges to standardise evaluation techniques, establish benchmarks, and track and evolve the state of the art. 
For example, the Blizzard Challenges have since their inception in 2005 (see \cite{black2005blizzard}) helped advance our sister field of text-to-speech (TTS) technology and identified important trends in the specific strengths and weaknesses in different speech-synthesis paradigms \cite{king2014measuring}. }

\moved{
In 2020 we organised the first gesture-generation challenge, the GENEA Challenge 2020 \cite{kucherenko2021large}. In 2022 we followed up with the second GENEA Challenge \cite{yoon2022genea}. In addition to being an exercise in benchmarking both new and previously-published gesture-generation methods, the results of that challenge have since helped improve gesture-generation benchmarking in other ways as well. Researchers have, for example, used the visualisation \cite{wang2021integrated}, and the objective \cite{bhattacharya2021speech2affectivegestures} and subjective \cite{yoon2021sgtoolkit} evaluation methodologies, as a basis for future research. The data has also been used to benchmark subsequent gesture-generation models \cite{ferstl2021expressgesture, yazdian2022gesture2vec}, and even for automatic quality assessment \cite{he2022automatic}. 
In this paper, we follow up by reporting on the third gesture-generation challenge, the GENEA Challenge 2023.
}
\end{commented}

\section{Evaluation setup}

The focus of the GENEA Challenge is on large-scale, crowdsourced subjective evaluation of the generated gestures through multiple user studies.
This year, we 
evaluated three aspects of the generated gestures:
\begin{description}
\item[Human-likeness:] Whether the motion of the virtual character looks like the motion of a real human, without explicitly considering the speech or any interlocutor behaviour. 
\item[Appropriateness for agent speech:] Whether the motion of the virtual character is appropriate for the given speech, controlling for the overall human-likeness of the motion.
\item[Appropriateness for interlocutor behaviour:] Whether the motion of the virtual character is appropriate for the given interlocutor behaviour (both speech and motion), controlling for the overall human-likeness of the motion and (as far as possible) for the effect of the agent's own speech.
\end{description}
We sometimes call the first two studies ``monadic'' and the third ``dyadic''.
Note that the models always had the interlocutor information as input, even in monadic evaluations.
All evaluations are built on the same avatar and visualisation software as used in last year's challenge (see \cite{kucherenko2023genea}).
More details about the three evaluations are given in \cref{sec:humlike,sec:approp,sec:dyadic}, respectively.

\subsection{Data}\label{ssec:data}
In order to evaluate gesture generation in a dyadic context, we supplement the recordings in last year's dataset \cite{yoon2022genea} with information from the conversation partner in the original Talking With Hands (TWH) data \cite{lee2019talking}. Specifically, we provide speech (audio and transcription), speaker ID, and motion for two parties; the \emph{main agent} (for which the task is to generate motion) and the \emph{interlocutor} (which is the other party in the conversation). As a minor improvement to last year's data, we set the height of the hips to the same for all speakers, since they all use the same skeleton. This makes the character's feet touch the ground in a consistent manner for all speakers. The dataset is publicly available at \href{https://doi.org/10.5281/zenodo.8199132}{doi.org/10.5281/zenodo.8199132}.

We provided three splits to the challenge participants: 1) a training split, composed of last year's training and validation data \cite{yoon2022genea}, 2) a validation split, equivalent to last year's test data, and 3) a test split, containing data that was completely left out from the 2022 challenge but with main-agent motion omitted. We augmented the training data by making each recorded conversation appear twice, with the conversational roles flipped. There was also an option for participating teams to leverage a version of the BEAT dataset \cite{liu2022beat}, retargeted and adapted to 
the challenge 
skeleton, but none of the submitting teams used it.

We created a \emph{core test set} consisting of 41 \emph{chunks} of approximately one minute each, restricted to recordings with finger motion tracking for the speaker chosen to be the main agent. We further extended the test set with 29 additional chunks in which the two sides of the conversation were mismatched, for use in the interlocutor awareness study (see \cref{ssec:dyadicdata} for details). Hence, the extended test set (core and additional test sets) consists of 70 chunks in total.

\subsection{Conditions evaluated}
\label{ssec:conditions}
In total, 12 teams participated in the GENEA Challenge 2023 evaluation.
The evaluation also included two baseline systems and natural motion taken from the motion-capture recordings from the speakers in the database.

We call each source of motion a \emph{condition}
rather than a ``system'' or ``model'', since 
the evaluation also includes avatar motion based on human mocap.
This natural motion was labelled \textbf{NA} (for ``natural'') in all evaluations.
We prefer not to use the term ``ground truth'', since there is no single, true way to move to a given speech and interlocutor.
The NA condition is intended as a top line, but in practice there can be some mocap artefacts and the motion is thus not always completely human-like, as seen from the stimuli and results of the 2022 challenge \cite{yoon2022genea}.

Sources of artificial, generated motion may be referred to as \emph{systems} or \emph{models}.
The 12 systems entered into the challenge by participating teams \cite{kim2023kuispl, yang2023diffusestylegest, harz2023feinz, zhao2023diffugesture, keti2023ANONYMOUSSTILL, windle2023uea, tonoli2023unicamp, deichler2023diffusion, schmuck2023kcl, chemburkar2023discrete, zhao2023gesture, korzun2023finemotion} are often called \emph{entries} or \emph{submissions}.

In addition to natural mocap and the different entries, the evaluations also included two baseline systems for automatic gesture-generation, both based on an entry from the GENEA Challenge 2022, specifically \citet{chang2022ivi}, which won the reproducibility award.
This means that a total of 15 conditions were included in the evaluation.
The two baselines were included with the intention to provide continuity and easier comparison between different years of the challenge, and also to track the progress of the field with respect to a fixed baseline.
(Note that none of the systems used as baselines in the 2020 evaluation were included this year, since those systems now are too far behind the state of the art to constitute reasonable points of comparison.)
The two baseline conditions were: 1) Monadic baseline (\textbf{BM}): The 2022 challenge entry from \citet{chang2022ivi} with slightly adjusted hyper-parameters, since the dataset is not the exact same. 2) Dyadic baseline (\textbf{BD}): An adaptation of \textbf{BM} that also takes information from the interlocutor into account as an extra input to the model.
\begin{commented}
\begin{description}
\item [Monadic baseline (BM)]
\mayberemove{
The 2022 challenge entry from \citet{chang2022ivi} with slightly adjusted hyper-parameters, since the dataset is not the exact same. 
}


\item [Dyadic baseline (BD)]
\mayberemove{
An adaptation of BM that also takes
information from the interlocutor 
into account, making the model dyadic. 
}
\end{description}
\end{commented}


In the results, each condition
was given a two-character label, a.k.a.\ condition ID.
The first character identifies the condition type, with `N' for motion-capture of natural human motion, `B' for a baseline, and `S' for a submission from a participating team.
The second character distinguishes between conditions within the same tier and type.
In particular, the challenge entries were labelled \textbf{SA}--\textbf{SL}.
These labels are anonymous and have no relationship to the team names or identities.


\subsection{Setup of the user studies}
\label{ssec:prolific}
Study participants (a.k.a.\ \emph{test takers}) were recruited through the crowdsourcing platform \href{https://www.prolific.co/}{Prolific}.
Participants were required to reside in a set of six English-speaking countries, specifically UK, IE, USA, CAN, AUS, and NZ.
A Prolific user could take any number of our studies, but could only participate in each study at most once.

\begin{figure}[!t]
\centering
   \includegraphics[width=\columnwidth]{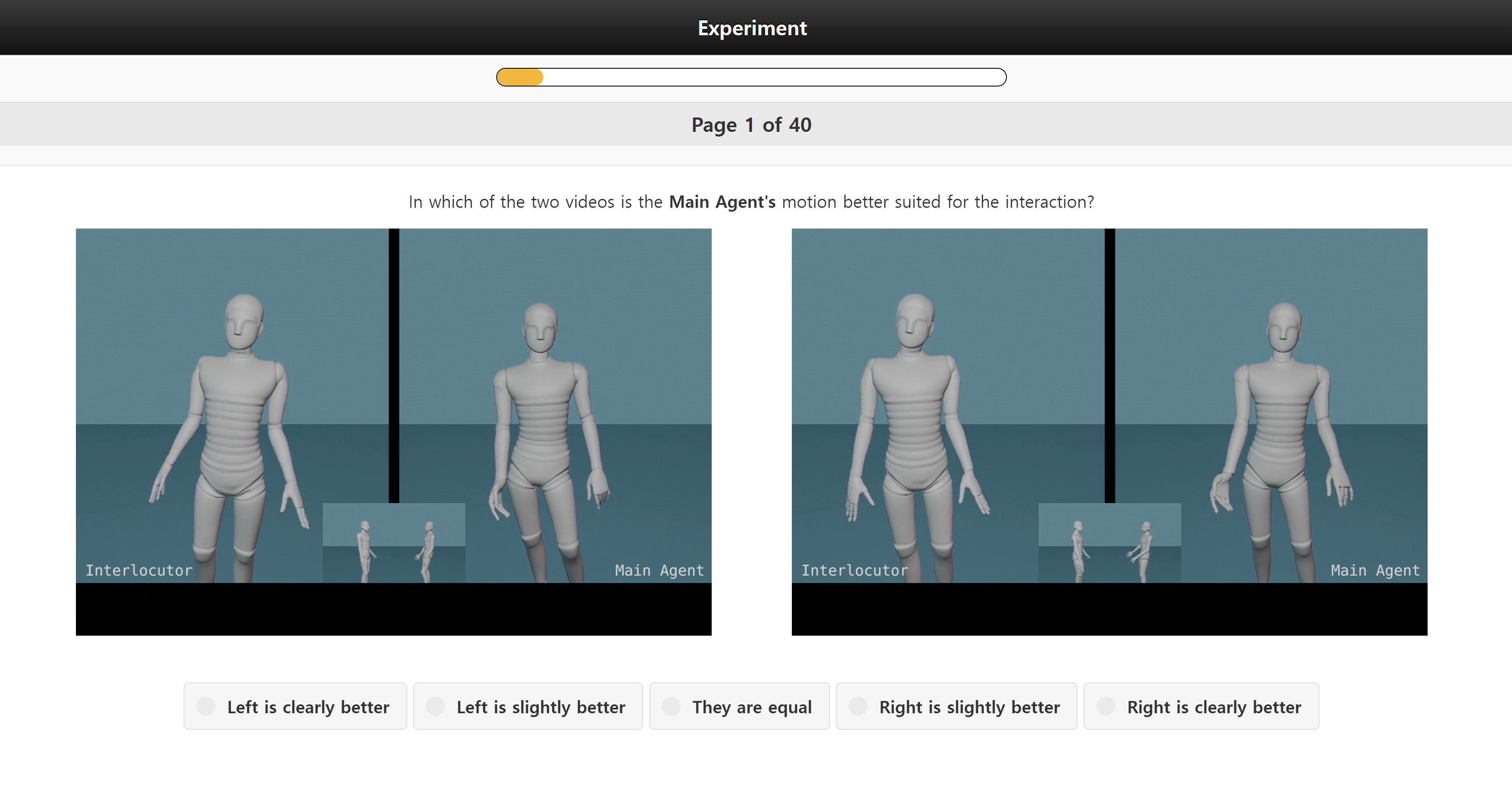}
\caption{Screenshots of the response interfaces from the appropriateness for the interlocutor evaluation, showing questions, response options, and snapshots of stimulus videos.}
\label{fig:appropgui}
\vspace{-2.6ex}
\end{figure}

While the specifics of the three user studies differed (see \cref{sec:humlike,sec:approp,sec:dyadic} for details), they also shared some common elements. In particular, we applied the same counterbalancing approach as in last year's challenge \cite{yoon2022genea}. Studies began with a screen of instructions, followed by a designated training page showing a fixed set of videos with different motions, to familiarise participants with the task.  Furthermore, after completing the study, participants filled in a short questionnaire to gather broad, anonymous
demographic information. We set the payment to 6 GBP for each study.

The studies also incorporated four attention checks per test taker similar to last year \cite{yoon2022genea}, to make sure that participants were paying attention to the task and to remove insincere test-takers. Each attention check was a modified stimulus that (partway through the stimulus) instructed the participant, using either a text overlay or a synthesised voice (only in studies where the audio was not muted), to select a specific response option.
Test takers who failed two or more attention checks were removed from the study without pay. 
You can find statistics about the test takers in \cref{tab:users}.

\subsection{Visualisation}
Like in previous years, we developed and maintained a standardised visualisation pipeline to which challenge participants had access.
This allowed participants to preview their generated motions and to roughly assess how the final stimuli videos during evaluation would look like.
We used the same virtual avatar (\cref{fig:appropgui}) in all rendered videos during the challenge and the evaluation.
The avatar was the same as last year \cite{yoon2022genea}.
The visualiser is capable of visualising both monadic and dyadic interactions; either one character or two characters are visible at a time.
All teams had access to the official visualisation and rendering pipeline
, in the form of code intended for local use with Blender.
The code is open source and is available at \href{https://github.com/TeoNikolov/genea_visualizer/}{github.com/TeoNikolov/genea\_visualizer/} and the rendered stimuli videos at {\href{https://doi.org/10.5281/zenodo.8211448}{doi.org/10.5281/zenodo.8211448}}.

\section{Human-likeness evaluation}
\label{sec:humlike}
The human-likeness evaluation for the GENEA Challenge 2023 used the HEMVIP methodology \cite{jonell2021hemvip}, same as in 2022, with some minor changes to GUI design, question formulation and response interface, and the statistical analysis.

\subsection{Data selection for stimuli}
\label{ssec:humsegments}
From the 41 chunks of core test set data, we selected 41 short segments of test speech and corresponding test motion to be used in the two agent-focussed subjective evaluations.
Our rules for selecting these segments were as follows:
\begin{enumerate*}[label=(\roman*)]
\item Segments should be around 8 to 10 seconds long.
\item Segments should only be taken from parts of the interaction
where the person chosen as the main agent is the active speaker (so no turn-taking within a segment, but backchannels from the interlocutor were OK).
\item Segments should not contain any parts where \citet{lee2019talking} had replaced the speech with silence for anonymisation. 
\item Segments should be more or less complete phrases, starting at the start of a word and ending at the end of a word. 
\item The recorded motion capture in the segments (i.e., the NA motion) should not contain any significant artefacts such as whole-body vibration.
\end{enumerate*}

The 41 segments selected in this way were between 7 and 13 seconds in duration, with the average duration being 9 seconds.

\subsection{Evaluation procedure}

On each page (a.k.a.\ screen) in the HEMVIP evaluation, eight different motion examples were presented in parallel \cite{jonell2021hemvip}, all corresponding to the same speech segment, but different conditions.


Each page asked participants ``Please indicate on a sliding scale how human-like the gesture motion appears''. Study participants gave their ratings in response to this question on a scale from 0 (worst) to 100 (best) by adjusting an individual GUI slider for each video.
In contrast to previous years, however, the numbers corresponding to the exact rating were not shown, and test takers were not explicitly told, e.g., that the sliders had 101 steps.
By using sliders without numerical labels, it is more clear to test takers that the scale is intended to be ordinal.

Like in \citet{kucherenko2021large,jonell2021hemvip}, the rating scale was anchored by partitioning the sliders into five equal-length intervals labelled (from best to worst) ``Excellent'', ``Good'', ``Fair'', ``Poor'', and ``Bad''.
These labels were based on those associated with the 5-point scale used in the Mean Opinion Score (MOS) \cite{itu1996telephone} standard for audio quality evaluation.
An example of the evaluation interface 
can be seen in the supplement. 

Since it has been found that speech content can influence gesture perception and confound motion evaluations \cite{jonell2020let}, the videos seen by participants in these human-likeness evaluations (although they all corresponded to the same speech input and had the same length) were completely silent and did not include any audio.
This way, ratings can only depend on the motion seen in the videos.

Each participant completed 10 pages of ratings for the evaluation.
The evaluation design was balanced in the same way as last year \cite{yoon2022genea} such that each segment appeared on pages 1 through 10 
with approximately equal frequency across all participants (segment order).
\toremove{
}




\begin{table}[!t]
\centering
\caption{Statistics on test takers and user-study responses.}
\label{tab:users}
\begin{tabular}{@{}l|c|c|c|c|c|c@{}}
\toprule 
 & No. & \multicolumn{2}{c|}{Videos per} & \multicolumn{2}{c|}{No.\ responses} & Median\tabularnewline
 & test & page & test & \multicolumn{2}{c|}{per condition} & time\tabularnewline
Study & takers &  & taker & min & max & taken\tabularnewline
\midrule
Human-likeness & 200 & 8 & 80 & \hphantom{0}938 & 2000 & 25.3 min\tabularnewline
Speech approp. & 600 & 2 & 80 & 1766 & 1815 & 23.8 min\tabularnewline
Interloc.\ approp. & 423 & 2 & 80 & \hphantom{0}993 & 1019 & 30.3 min\tabularnewline
\bottomrule
\end{tabular}
\end{table}

\subsection{Response data}
\begin{figure}[t!]
\centering%
\includegraphics[width=\columnwidth]{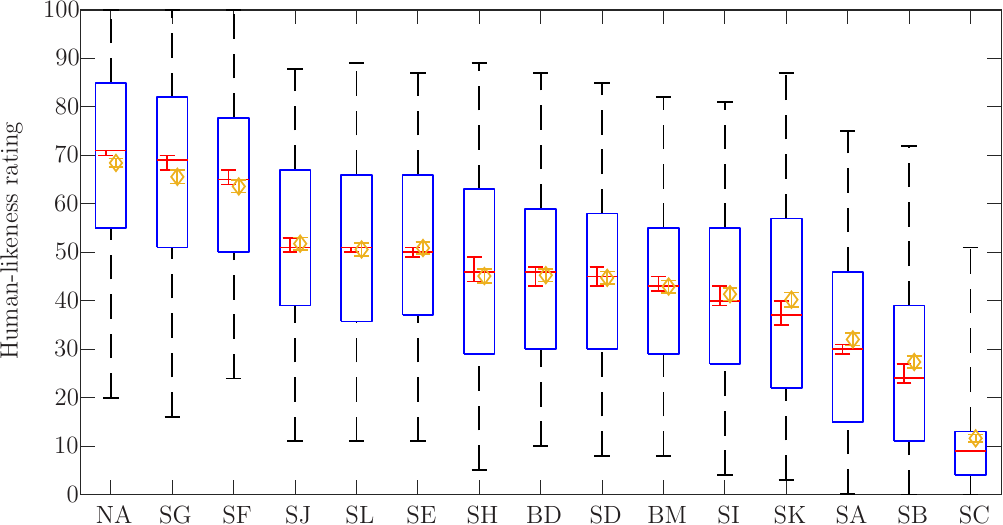}
\caption{Box plot visualising the rating distribution in the human-likeness study. Red bars are the median ratings (each with a 0.05 confidence interval); yellow diamonds are mean ratings (also with a 0.05 confidence interval). Box edges are at 25 and 75 percentiles, whilst whiskers cover 95\% of all ratings for each condition. Conditions are ordered by descending sample median rating.}
\label{fig:humlikeboxplot}
\end{figure}%
Confidence intervals for the median were computed using order statistics, leveraging the binomial distribution cumulative distribution function, cf.\ \cite{hahn1991statistical}, while those for the mean used a Gaussian assumption (i.e., using Student's $t$-distribution cdf, rounded outward to ensure sufficient coverage).

The rating distribution in the study is further visualised through the box plot in \cref{fig:humlikeboxplot}.
The rating distributions in the figure are seen to be quite broad.
This is common in evaluations like HEMVIP, since the range of the responses not only reflects differences between conditions, but also extraneous variation, e.g., between stimuli, in individual preferences, and in how critical different raters are in their judgements.
In contrast, the plotted confidence intervals are seen to be quite narrow, due to the large number of ratings collected for each condition.
\begin{figure}[t!]
\centering%
\includegraphics[width=\columnwidth]{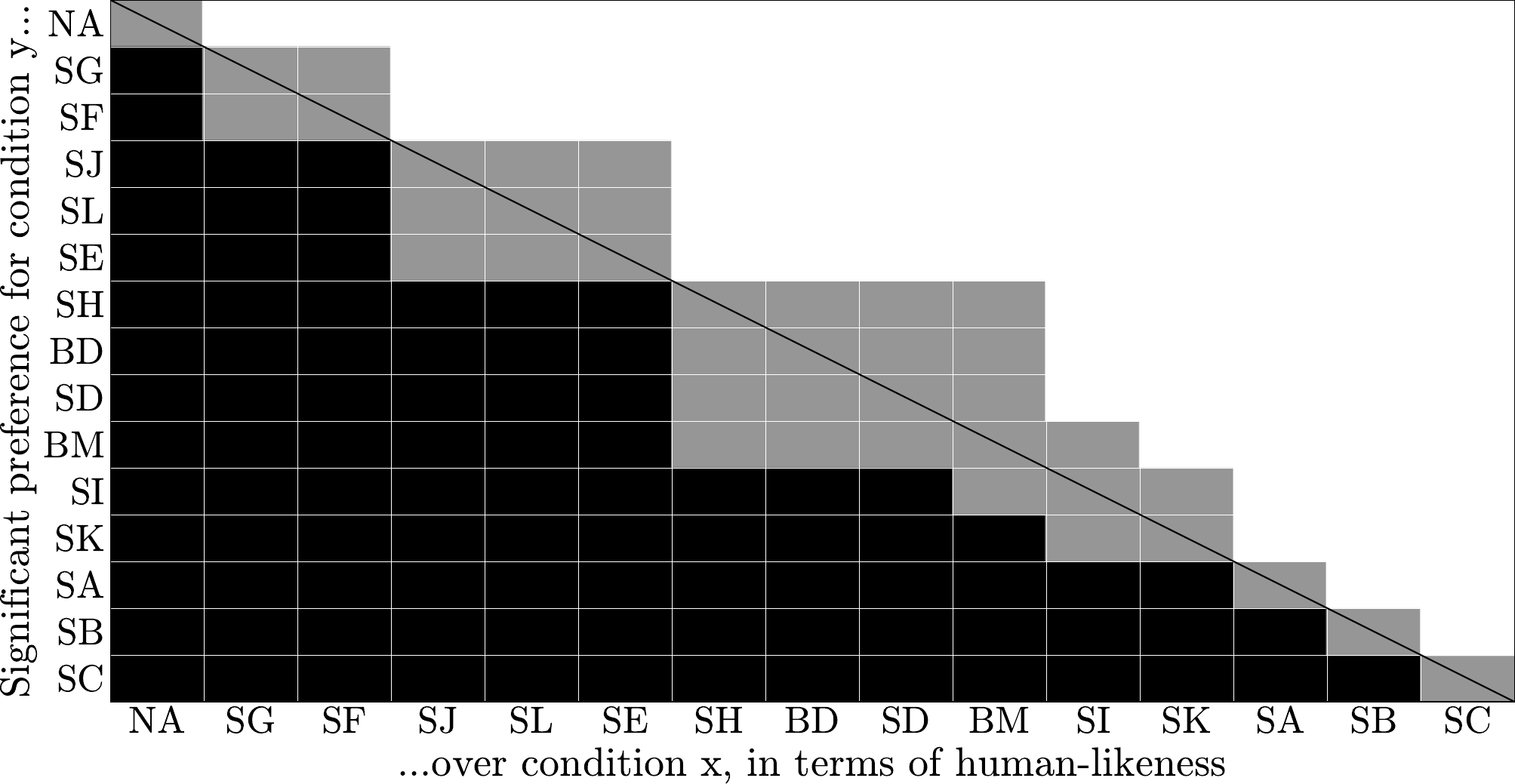}
\caption{Significance of pairwise differences between conditions in human-likeness. White means that the condition listed on the $y$-axis rated significantly above the condition on the $x$-axis, black means the opposite, and grey means no statistically significant difference at the level $\alpha=0.05$ after Holm-Bonferroni correction. Conditions are listed in the same order as in \cref{fig:humlikeboxplot}.}
\label{fig:humlikedifferences}
\vspace{-1.2em}
\end{figure}

\subsection{Significant differences}
\label{ssec:pairwisehumlike}
To analyse the significance of differences in median rating between different conditions, we applied two-sided Mann-Whitney $U$ tests to all unordered pairs of distinct conditions in each study.
This is an unpaired nonparametric test, which differs from the pairwise Wilcoxon signed-rank tests used in previous GENEA Challenges.
The reason for this change is that, whilst pairwise tests can more easily control for variation between segments and raters, such tests only can be used for data from screens where both conditions under comparisons were shown to participants.
Since this year's challenge is comparing more conditions than before but still uses the same number of sliders (8 in parallel), such pairings are relatively rare.
An unpaired test can use all the available ratings of any given condition, making it more powerful in this instance.

Unlike Student's $t$-test, which assumes that the test statistic follows a Gaussian distribution, this analysis is valid also for ordinal response scales, like those we have here.
For each condition pair, only pages for which both conditions were assigned valid ratings were included in the analysis of significant differences.


The $p$-values computed in the significance tests were adjusted for multiple comparisons using the Holm-Bonferroni method \cite{holm1979simple}, which is uniformly more powerful than conventional Bonferroni correction, to keep the family-wise error rate (FWER) at or below $\alpha=0.05$.
After FWER correction, our statistical analysis found all but 12 out of 105 condition pairs to be significantly different at the level $\alpha=0.05$.
Which conditions that were found to be rated significantly above or below which other conditions in the study is shown in \cref{fig:humlikedifferences}.

\subsection{Discussion}
\label{ssec:humlikecomments}
Looking at the results, we see that the challenge submissions span a wide range of different human-likeness scores, with a greater concentration near the middle.
Unlike the GENEA Challenge 2022 \cite{yoon2022genea}, no submission achieved median human-likeness ratings that exceeded those of the human motion capture, although two submissions were not far behind.
This helps establish the trend that, whilst human-like gesture motion is not altogether a solved problem, strong performance in that regard is possible (but not necessarily easy) with current data-driven methods.
The baselines BD and BM performed similarly to each other and achieved ratings near the middle of the pack.
This resembles the performance of the underlying system last year \cite{chang2022ivi,kucherenko2023evaluating}, when it was a challenge submission.

Similar to past years' challenges, natural mocap (NA) is rated significantly lower than perfect human-likeness.
Among other things, this may be attributed to shortcomings with the mocap, the 3D model used (which deliberately appears neutral and lacks mouth and gaze), and human biases against assigning high scores.

One other noteworthy observation to keep in mind for later is the fact that ordering conditions by their median human-likeness (as in \cref{fig:humlikeboxplot}) produces a quite different ordering compared to ordering systems by appropriateness in \cref{sec:approp,sec:dyadic}.
This indicates that we managed to disentangle human-likeness and appropriateness in our evaluation.%
\begin{table*}[!t]
\caption{\label{tab:stimuli} Overview of the stimuli used in the different studies.
Numbers 1 and 2 indicate which of two test-set segments -- the matched or the mismatched -- that data was sourced from. 
Each matched segment was paired with a corresponding mismatched segment elsewhere in the test data. 
Abbreviations: ``Hum.-like.'' stands for ``Human-likeness'', ``approp.'' stands for ``appropriateness'', ``segs.'' stands for ``segments'', ``ID'' stands for ``Speaker ID'', and ``Interloc.'' for ``Interlocutor''.
The motion for condition NA was taken from segment 1 in matched stimuli and 2 in mismatched. All other conditions used motion synthesised from the inputs listed under ``Source of test-time system inputs''.
}
\begin{tabular}{@{}l|ccc|c|cc|ccc|c|cc@{}}
\toprule
 & \multicolumn{3}{c|}{Studies featured} & Active & \multicolumn{5}{c|}{Source of test-time system inputs} & \multicolumn{3}{c}{Source of what is presented in stimuli}\tabularnewline
 & Hum.- & Speech & Dyadic & speaker & \multicolumn{2}{c|}{Main agent } & \multicolumn{3}{c|}{Interlocutor} & Main agent & \multicolumn{2}{c}{Interlocutor}\tabularnewline
Stimulus type & like. & approp. & approp. & in segs. & Speech & ID & Speech & Motion & ID & speech & Speech & Motion\tabularnewline
\midrule
Matched (speaking) & \cmark & \cmark &  & Agent & 1 & 1 & 1 & 1 & 1 & 1 & \multicolumn{2}{c}{N/A (monadic)}\tabularnewline
Mismatched (speaking) &  & \cmark &  & Agent & 2 & 2 & 2 & 2 & 2 & 1 & \multicolumn{2}{c}{N/A (monadic)}\tabularnewline
\midrule
Matched (listening) &  &  & \cmark & Interloc. & 1 & 1 & 1 & 1 & 1 & 1 & 1 & 1\tabularnewline
Mismatched (listening) &  &  & \cmark & Interloc. & 1 & 1 & 2 & 2 & 2 & 1 & 1 & 1\tabularnewline
\bottomrule
\end{tabular}
\end{table*}

\section{Appropriateness for agent speech}
\label{sec:approp}
Similar to the GENEA Challenge 2022 \cite{yoon2022genea}, we also evaluated the appropriateness of the generated motion for the main agent's speech, using a similar but improved methodology compared to 2022.

The GENEA Challenge 2022 adapted a new mismatching methodology \cite{jonell2020let,rebol2021passing} with the design goal to evaluate motion appropriateness whilst controlling for the human-likeness of the motion in an effective way.
Test takers were presented with a pair of videos, each video showing motion corresponding to a different speech recording. We replaced the audio track of one of the videos with the speech audio of the other clip, resulting in one video where the motion matched the audio, and another video where audio and motion were from unrelated sources. 
Participants were then asked to pick one video from the pair 
where motion, according to them, best matched the speech.
Their responses reveal how specific the gesture motion is to the given speech. 

Since last year's appropriateness evaluation, with only three possible response options, had quite wide confidence intervals, we asked test takers to indicate their degree of preference this year, so as to get information value out of each response.
This change was based on a suggestion for future evaluations in \citet{kucherenko2023evaluating} and piloted in \citet{mehta2023diffttsg}.
More specifically, test takers were asked ``Which character's motion matches the speech better, both in terms of rhythm and intonation and in terms of meaning?''
The five response options 
were adapted from \cite{mehta2023diffttsg}, merely changing ``much'' to ``clearly''. Which of the two videos (left or right) was the matched one was random.
A screenshot of the evaluation interface used for the appropriateness studies is presented in the supplement. 

In this type of evaluation, a system whose output does not depend on the input will perform at chance rate.
This avoids the issues seen in the appropriateness evaluation of the GENEA Challenge 2020 \cite{kucherenko2021large}, and we no longer recommend that earlier methodology.


\subsection{Stimuli}
\cref{tab:stimuli} provides an overview of the different stimuli used in the different user studies.
The speech appropriateness evaluation leveraged the same evaluation segments chosen for the human-likeness evaluation in \cref{ssec:humsegments}.
Concretely, we created the mismatched stimuli by taking the speech and motion for the 41 segments selected there, and then permuting the motion in between them such that no motion segment ever remained in its original place.
Since the different segments generally do not have the same length, sometimes a longer or shorter segment of motion had to be excerpted from the test-set chunks (original or generated), so as to match the new speech duration, but the starting point of the motion video was always the same as in the respective matched stimulus video (i.e., corresponding to the start of a phrase).
In the terminology of \cref{tab:stimuli}, the mismatched motion segments were excerpted using the start time of segment 2 but the duration of segment 1.
This is exactly the same as was done in 2022.


\subsection{Evaluation interface and question asked}
\label{ssec:appropui}
Because we are deliberately avoiding comparisons across conditions (as different conditions may have different motion qualities), test takers were only asked to view and compare a single pair of videos at a time.
For pairwise comparisons like these, asking for a preference between two items can be more efficient than asking participants to rate each video 
\cite{wolfert2021rate}.
We therefore asked for preference rather than absolute ratings in these studies. The question asked was "Which character's motion matches the speech better, both in terms of rhythm and intonation and in terms of meaning?".

\subsection{Response data}

For analysis, the five possible responses were converted to integer values $\{-2,\,-1,\,0,\,1,\,2\}$ in order, with $-$2 meaning the mismatched stimulus was rated as clearly better and 2 meaning the matched stimulus was rated as clearly better.
Raw response statistics for all conditions in each of the two studies are available in the appendix.

\begin{table}[!t]
\centering%
\caption{Mean appropriateness scores (MAS) from both appropriateness studies with confidence intervals at the level $\alpha=0.05$. Submissions are ordered alphabetically.}
\label{tab:approp}
\begin{tabular}{@{}lcc|lcc@{}}
\toprule 
Con- & \multicolumn{2}{c|}{MAS for\ldots} & Con- & \multicolumn{2}{c}{MAS for\ldots} \\
dition & Speech & Interloc. & dition & Speech & Interloc. \\
\midrule
NA & \hphantom{$-$}0.81$\pm$0.06 & \hphantom{$-$}0.63$\pm$0.08 & SF & \hphantom{$-$}0.20$\pm$0.06 & \hphantom{$-$}0.04$\pm$0.06 \\
BD & \hphantom{$-$}0.14$\pm$0.06 & \hphantom{$-$}0.07$\pm$0.06 & SG & \hphantom{$-$}0.39$\pm$0.07 & $-$0.09$\pm$0.08 \\
BM & \hphantom{$-$}0.20$\pm$0.05 & $-$0.01$\pm$0.06 & SH & \hphantom{$-$}0.09$\pm$0.07 & $-$0.21$\pm$0.07 \\
SA & \hphantom{$-$}0.11$\pm$0.06 & \hphantom{$-$}0.09$\pm$0.06 & SI & \hphantom{$-$}0.16$\pm$0.06 & \hphantom{$-$}0.04$\pm$0.08 \\
SB & \hphantom{$-$}0.13$\pm$0.06 & \hphantom{$-$}0.07$\pm$0.08 & SJ & \hphantom{$-$}0.27$\pm$0.06 & $-$0.03$\pm$0.05 \\
SC & $-$0.02$\pm$0.04 & $-$0.03$\pm$0.05 & SK & \hphantom{$-$}0.18$\pm$0.06 & $-$0.06$\pm$0.09 \\
SD & \hphantom{$-$}0.14$\pm$0.06 & \hphantom{$-$}0.02$\pm$0.07 & SL & \hphantom{$-$}0.05$\pm$0.05 & \hphantom{$-$}0.07$\pm$0.06 \\
SE & \hphantom{$-$}0.16$\pm$0.05 & \hphantom{$-$}0.05$\pm$0.07 & & & \\
\bottomrule
\end{tabular}
\end{table}

The table also includes 95\% confidence intervals for the average numerical value of the per-condition user responses, computed using a Student's $t$-distribution.
We refer to this average, used for our appropriateness statistical analyses, as \emph{mean appropriateness score} (or \emph{MAS}).
\begin{figure*}[t!]
\centering%
  \begin{subfigure}[b]{\columnwidth}
    \centering%
    \includegraphics[width=\columnwidth]{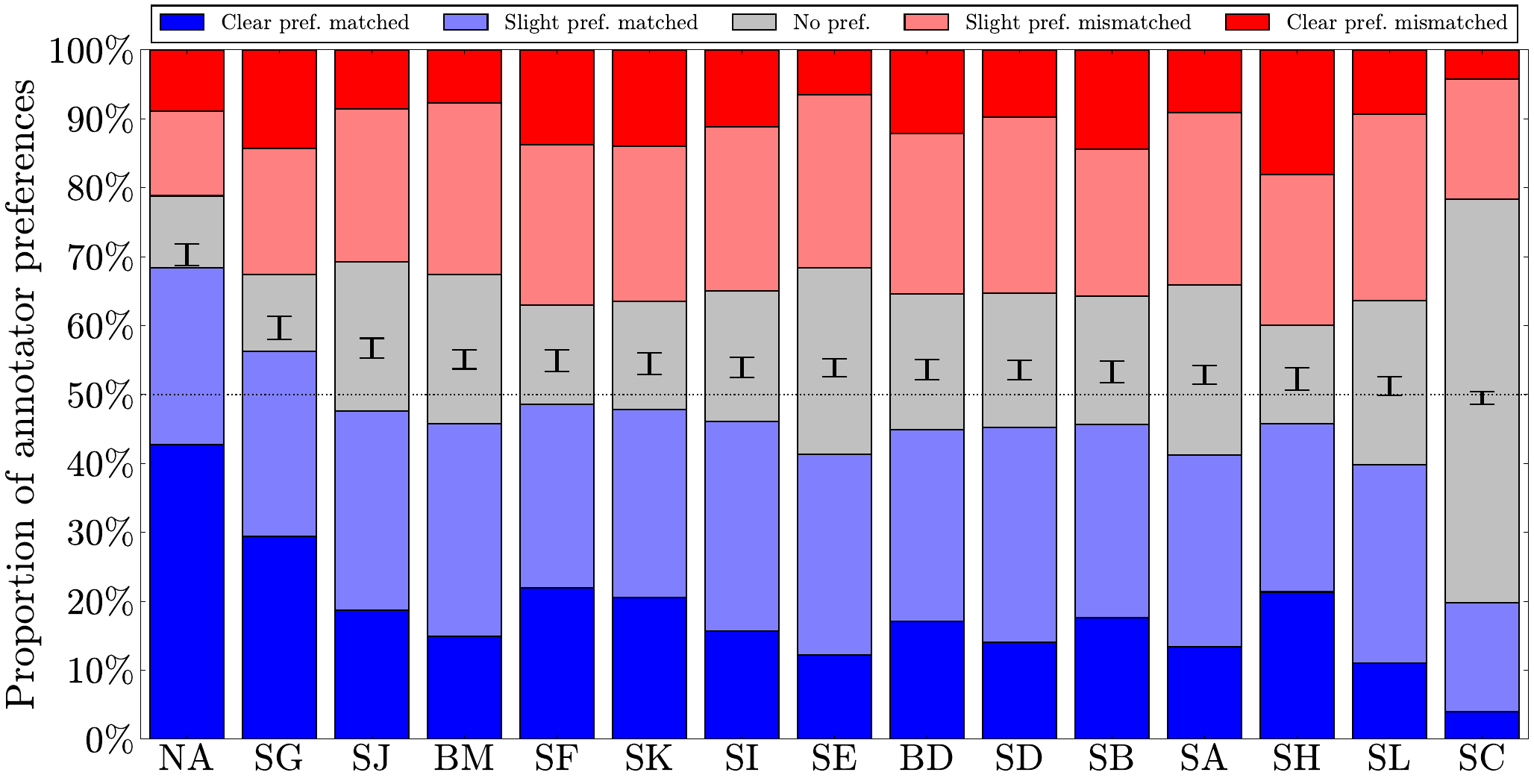}
    \caption{Appropriateness for agent speech}
    \label{sfig:monadicappropbars}
  \end{subfigure}
\hfill
  \begin{subfigure}[b]{\columnwidth}
    \centering%
    \includegraphics[width=\columnwidth]{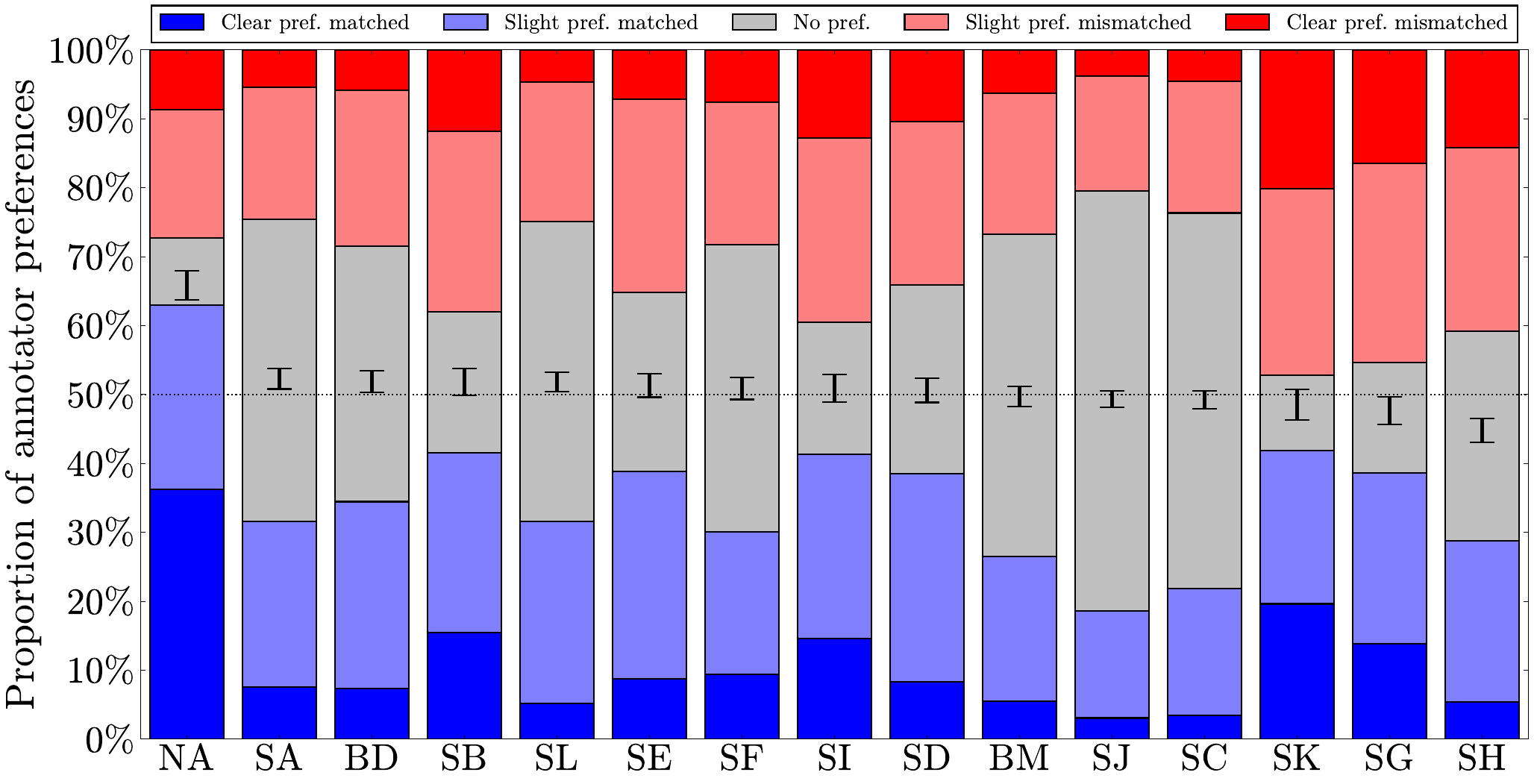}
    \caption{Appropriateness for the interlocutor}
    \label{sfig:dyadicappropbars}
  \end{subfigure}
\caption{Bar plots visualising the response distribution in the appropriateness studies. 
On top of each bar is also a confidence interval for the mean appropriateness score, scaled to fit the current axes.
The dotted black line indicates chance-level performance. Conditions are ordered by mean appropriateness score.} %
\label{fig:appropbars}
\end{figure*}

The response distributions in the study are further visualised through the bar plot in  \cref{sfig:monadicappropbars}.
The bar plot also visualises the confidence interval for the mean appropriateness score, but scaled linearly from the interval $[-2,\,2]$ to [0\%,\,100\%] in order to make the intervals comparable with the rest of the plot.

\subsection{Statistical analysis}
\label{ssec:monadicanalyses}
Two different statistical analyses were applied to the results, testing:
\begin{enumerate}
\item Whether the mean appropriateness score of a given condition is statistically different from chance performance (without correcting for multiple comparisons).
\item Whether the mean appropriateness score of any given condition was statistically different from the mean appropriateness score (correcting for multiple comparisons).
\end{enumerate}
All statistical tests were carried out at the $\alpha=0.05$ level.

Remember that in our evaluation, a system whose output does not depend on the input will theoretically perform at a chance rate.

\subsubsection{MAS differences from chance performance}
The dotted line at 50\% (equal to an MAS of zero) in \cref{sfig:monadicappropbars} marks chance performance.
Any condition whose 95\% MAS confidence interval does not overlap with zero
demonstrates performance that is statistically different from chance (without any correction for multiple comparisons).
On this measure, all conditions except SL and SC were statistically better than chance in terms of MAS.

\subsubsection{MAS differences between conditions}
To assess whether any two conditions were statistically different from one another, we used Welch's $t$-test.
This is an unpaired statistical test; the study design was not optimised for pairwise testing since a re-analysis of the data from \citet{mehta2023diffttsg} to compare the effect of paired versus unpaired $t$-tests for analysis
produced no noteworthy differences.


A correction for multiple comparisons was performed to control the false discovery rate (FDR), i.e., the fraction of erroneous rejections of the null hypothesis out of all rejected null hypotheses.
Specifically, we used the BH non-adaptive one-stage linear step-up procedure \cite{benjamini1995controlling} as implemented in \href{https://puolival.github.io/multipy/}{the MultiPy package}.
Note that controlling the FDR is not the same as controlling for the familywise error rate (which controls the probability of making any false rejections at all), as done in \cref{sec:humlike}.

\begin{figure*}[t!]
\centering%
  \begin{subfigure}[b]{0.49\textwidth}
    \centering%
    \includegraphics[width=\textwidth]{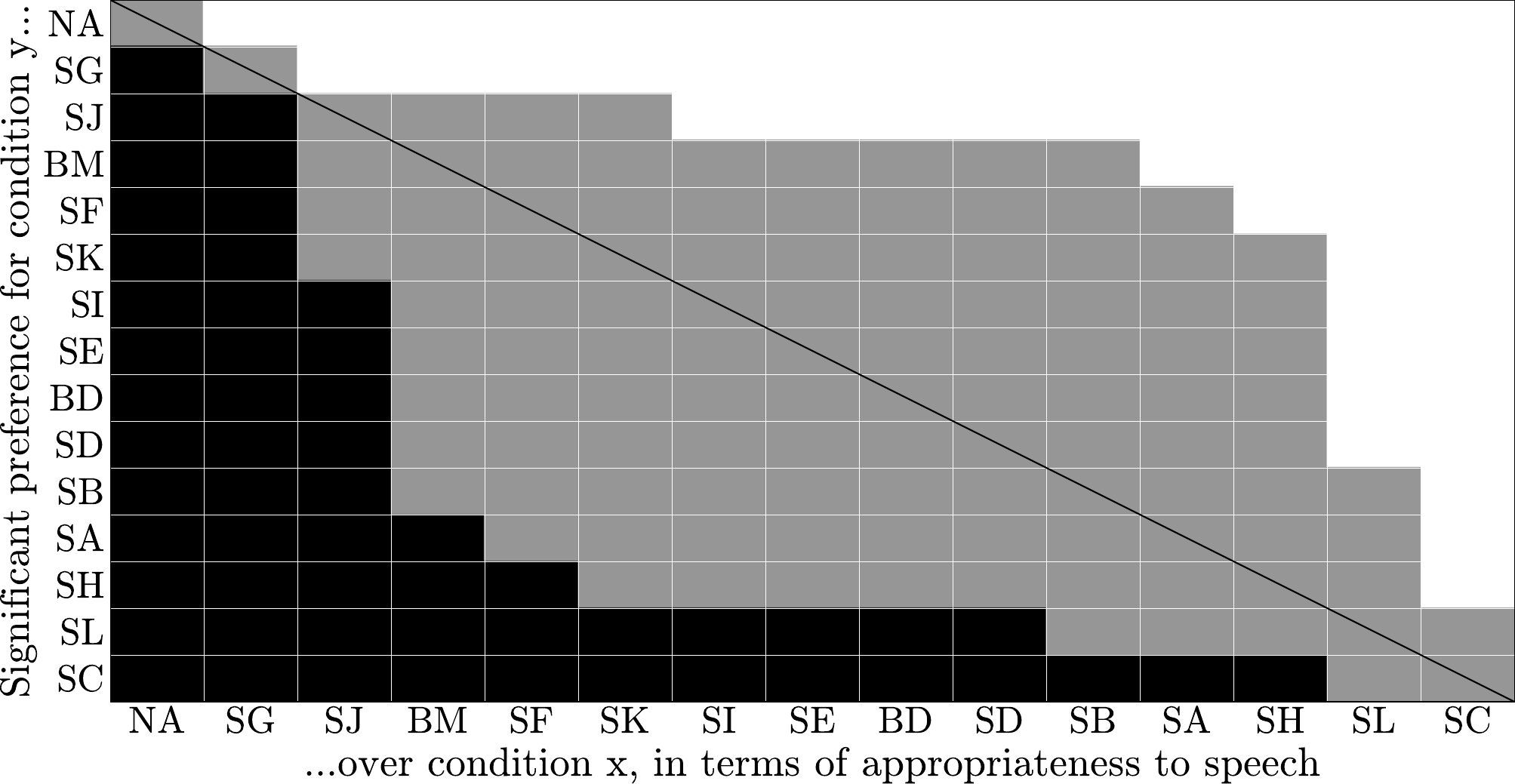}
    \caption{Appropriateness for agent speech}
    \label{sfig:monadicdifferences}
  \end{subfigure}
  \begin{subfigure}[b]{0.49\textwidth}
    \centering%
    \includegraphics[width=\textwidth]{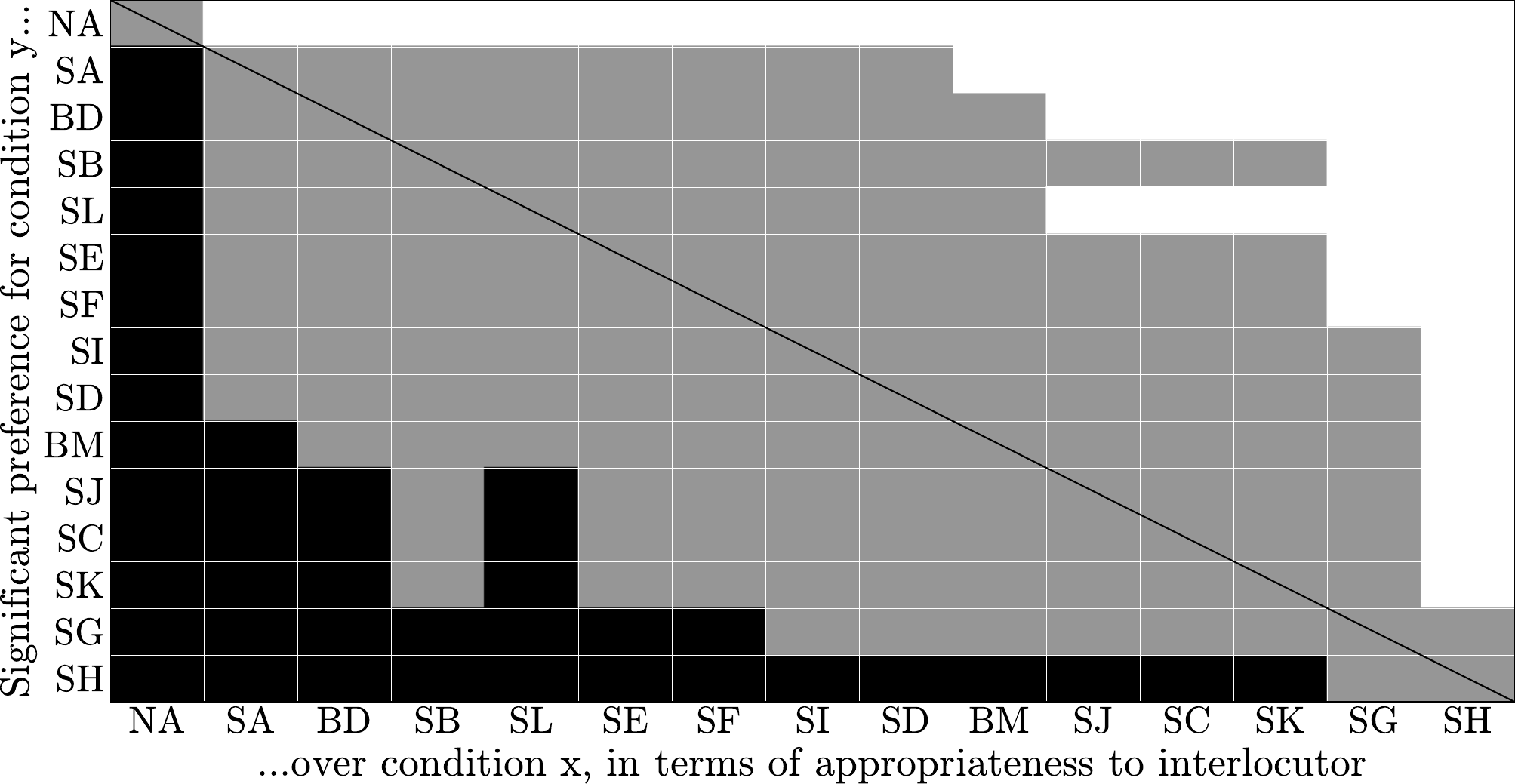}
    \caption{Appropriateness for the interlocutor}
    \label{sfig:dyadicdifferences}
  \end{subfigure}
\caption{Significant differences between conditions in the two appropriateness studies. White means the condition listed on the $y$-axis achieved an MAS significantly above the condition on the $x$-axis, black means the opposite 
, and grey means no statistically significant difference at level $\alpha=0.05$ after correction for the false discovery rate. Conditions use the same order as the corresponding subfigure in \cref{fig:appropbars}.}
\label{fig:appropdifferences}
\end{figure*}

After FDR correction, our statistical analysis found 56 of 105 condition pairs to be significantly different. \cref{sfig:monadicdifferences} visualises the statistically significant differences between conditions.


\subsection{Discussion}
Just as in the GENEA Challenge 2022 results, natural motion capture (condition NA) exhibited the greatest difference between matched and mismatched stimuli.
NA attained an MAS of $0.81\pm0.06$ on the scale from $-$2 to 2.
A more in-depth discussion of why this magnitude of difference is reasonable (and why one should not expect 100\%) can be found in \cite{kucherenko2023evaluating}.

Unlike the human-likeness evaluation, however, no submission came close to NA in terms of specific appropriateness to the speech in this evaluation, with the best synthesis system achieving a MAS around 0.39 and a 62\% preference for matched motion after splitting ties.
Also unlike the case for human-likeness, most submissions are confined to a narrow range of MAS scores between 0.27 and 0.10, or so.
Clearly, current synthesis methods are far from as successful as humans are in producing gestures that match the specific speech, although the current evaluation does not reveal to what extent the shortcomings seen are related to mistimed gesturing, or to failures to generate semantically relevant gesticulation.
Either way, it is not clear if relatively non-specific gesture motion, as generally produced by the systems in the challenge, enhances human-agent communication.
Since better communication is a major motivator of research into (and applications of) embodied conversational agents in the first place, our results suggest that improving the appropriateness of synthesised gestures for agent speech is an important research target.

\section{Appropriateness for the interlocutor}
\label{sec:dyadic}
Although there were many similarities to the appropriateness evaluation in \cref{sec:approp}, this study differed in several important aspects, namely in what was being studied, how the motion contained in the stimuli was obtained, and how the stimuli were presented for evaluation.


For this final evaluation, we wanted to use mismatching 
to quantify motion appropriateness for the interaction 
whilst controlling for both motion human-likeness \emph{and} for the effect of the main agent's own speech as far as possible.
This required carefully crafted stimuli.

\subsection{Data creation for stimuli}
\label{ssec:dyadicdata}
We reasoned that interlocutor-aware behaviour might be most salient when the interlocutor is the active speaker, and the main agent is reacting to them.
For this reason, this appropriateness evaluation exclusively considered segments where the interlocutor was the active speaker. 

Prior to the test set being shared with challenge teams, we selected 29 segments from the core test set where the main agent exhibited dyadic adaptations to the interlocutor, for example head nods.
The segments selected in this way were between 5.5 and 15 seconds in duration, with the average duration being 10.4 seconds.

For each segment that we selected from the test set, we proceeded to create a second version of the same interaction, where the main agent was the same (same WAV, TSV, and speaker ID) but the interlocutor was replaced by mismatched interlocutor behaviour (WAV, TSV, and speaker ID) taken from another interaction.
The mismatched interlocutor behaviour was chosen such that the interlocutor was the active speaker for the duration of the segment used. 
These mismatched interactions, one interaction for each segment used for the present study, formed the extended test set mentioned in \cref{ssec:data}.
When participating teams generated motion for the extended test set, they would thus create motion examples based on the same main-agent speech as the matched segments, but using different interlocutor behaviour as input.
This can be used to create stimulus pairs where any systematic differences in main agent motion is attributable only to whether or not the motion was generated for the matched interlocutor behaviour or not.
See \cref{tab:stimuli} for a complementary specification of this setup.



\subsection{Stimuli}
\begin{commented}
\mayberemove{Unlike all other evaluations, stimuli for the interlocutor-appropriateness evaluation were dyadic -- they presented both the main agent and interlocutor behaviour together, so that test takers would be able to see and assess the interaction between the two parties.}
\end{commented}
\cref{fig:appropgui} shows static images of example video stimuli that contains two frontal views, one of each party in the conversation, and a third view showing the two interacting characters from the side.

Unlike the stimuli in previous appropriateness studies, the audio in the stimuli video clips was presented in stereo, with the interlocutor audio in the left channel and the main agent audio in the right channel.
This made it possible to tell who was who and who was saying what despite the two avatars not having mouths. 


Stimuli (whether matched or mismatched) would only use audio from the matched segment/interaction.
Similarly, the interlocutor motion would also be taken from the matched interaction for both matched and mismatched stimuli.
Thus the only difference between the two video stimuli in a pair would be in how the main agent moves.
In the matched case, it would be based on the speech and the interlocutor motion seen in the video, but in the mismatched case, it would be based on the same agent speech but different interlocutor speech and motion, leveraging the artificially-constructed extended test set.
The only exception would be for condition NA, where the mismatched main-agent motion would, by necessity, be based on mismatched rather than matched agent speech.

\subsection{Evaluation interface and question asked}
The evaluation interface for the user study of motion appropriateness for the interlocutor was exactly the same as in \cref{ssec:appropui}.
This time, test takers were asked ``In which of the two videos is the Main Agent's motion better suited for the interaction?''
A screenshot of the complete evaluation interface can be seen in \cref{fig:appropgui}.

After an instructions page and a training page
, each study participant evaluated 40 pages (i.e., 40
 pairs of videos) in a counterbalanced design.
This means that each person watched 80 videos in total.
A design goal of this study was that each condition would receive approximately 1,000 or more responses, similar to before.

\begin{commented}
\subsection{Attention checks and study participants}

\mayberemove{The instructions on Prolific listed stereo audio as a strict requirement for taking the test, since stereo was necessary for telling main agent and interlocutor speech apart.
To ensure compliance with this requirement, the entire test was preceded by a page for checking that test takers indeed had (correctly lateralised) stereo audio.}
\end{commented}


Despite the same number of videos watched, the time taken per participant was greater than in \cref{sec:approp}. One explanation might be that the task is more demanding, with multiple characters to pay attention to.

\subsection{Response data}

The statistical analysis used mean appropriateness scores based on the integer values $\{-2,\,-1,\,0,\,1,\,2\}$ as before.
Raw response statistics for all conditions in each of the two studies are shown in \cref{tab:approp}, along with 95\% confidence intervals for the MAS, all computed as before.

Like before, the response distributions from the study are further visualised through the bar plot in \cref{sfig:dyadicappropbars}, together with confidence intervals for the mean appropriateness score, scaled to fit the rest of the plot.

\subsection{Statistical analysis}
\label{ssec:dyadicanalyses}
The same two statistical analyses as in \cref{ssec:monadicanalyses} were applied. 

\subsubsection{MAS differences from chance performance}
The dotted line at 50\% in \cref{sfig:dyadicappropbars} marks chance performance.
Any condition whose 95\% MAS confidence interval does not overlap with zero (which coincides with the dotted line in the graph) demonstrates performance that is statistically different from chance (without any correction for multiple comparisons).
On this measure, NA, BD, SA, and SL are significantly better than chance, whilst SG and SH instead are significantly worse.
The remaining conditions were not statistically different from chance.

\subsubsection{MAS differences between conditions}
To assess whether any two conditions were statistically different from one another, we once again used Welch's $t$-test together with the BH non-adaptive one-stage linear step-up procedure \cite{benjamini1995controlling} to control the false discovery rate.
After FDR correction, our statistical analysis found 45 of 105 condition pairs to be significantly different.
\cref{sfig:dyadicdifferences} visualises these statistically significant differences
using the same condition order as the box plot.

\subsection{Discussion}
To begin with, it must be pointed out that condition NA in this study is not comparable to the others, since its motion is mismatched not only with respect to the interlocutor, but also with respect to the main agent's own speech.
This cannot be avoided, since the interactions in the extended test set 
do not have any counterparts in real interactions.
The performance of NAs is thus in a sense a (likely unattainable) upper bound on MAS in this experiment.
However, it does establish that mismatched behaviour can reliably be detected by the test takers, even for cases where the main agent is listening rather than speaking, albeit with a slightly lower MAS than for the speaking segments in \cref{tab:approp}.

In general, we may expect the difference between matched and mismatched stimuli to be smaller in the dyadic study, since for the monadic study, all inputs to the models differed when generating the two motions being compared, whereas only the model inputs corresponding to the interlocutor differed for this study.
The greater fraction of ``They are equal'' responses in \cref{sfig:dyadicappropbars} compared to \cref{sfig:monadicappropbars}
indicates that this is indeed the case.

Having 5 out of 14 artificial systems differ from chance performance at level $\alpha=0.05$ is very unlikely to occur by random happenstance despite the absence of a correction for multiple comparisons.
This strongly suggests that a notable fraction of the dyadic systems indeed create meaningfully different motion depending on the behaviour of the interlocutor, although the effect in absolute terms is small.


Looking at individual conditions, we can make some additional observations.
Unlike previous studies, the two baselines are now far apart in the rankings, with BD (as hoped) being significantly different from chance performance, whilst BM (as expected) was not.
The MAS of conditions SG and SH are also notable.
Condition SG has gone from one of the most appropriate conditions (for the agent speech) to one of the least appropriate (for the interlocutor/interaction) in \cref{tab:approp}.
Condition SH, meanwhile, exhibited the largest effect size of any artificial system in this dyadic evaluation, but in the direction of being \emph{less} appropriate than random chance.
It is not clear to what extent this may be influenced by to the use of listening rather than speaking, or by the ``chimeric'' and artificial inputs in the extended test set, or even by the use of a dyadic visualisation.

\section{Conclusion}
We have hosted the GENEA Challenge 2023 to investigate and evaluate different techniques of speech-driven gesture-generation.
\begin{commented}
\mayberemove{
With a common ground setting
, we studied how each method performs in terms of rendering agent behaviours that are human-like, appropriate to its speech and appropriate to its interlocutor.
}
\end{commented}
Via our evaluation, we remark that it is challenging to generate human-like and speech-appropriate gestures that also address the problem of assuring appropriateness with respect to the behaviour of the interlocutor. 
\begin{commented}
\mayberemove{
This observation is seen with the submitted systems that perform well in terms of human-likeliness and appropriateness for agent speech but not for the appropriateness for the interlocutor. The performance of the systems varies over a large spectrum and a few of them are close to the human mocap for human-likeliness or distinctively better than chance for appropriateness.}
\end{commented}
Nevertheless, it is promising to see that a substantial amount of submissions managed to respond to the interlocutor. 
Notably, from the dyadic systems that consider the information from the interlocutor, 
we remark that the interlocutor’s information surely plays a role in rendering reactive motions that match the interlocutor’s behaviours.
Communication is not only mono-directional or bi-directional, but also multi-party. We envision observing more research conducted on dyadic interaction in the short-term and on multi-party interaction in the mid-term future to address all types of interaction.
ECAs use synthetic speech to render their utterances.
Recent studies show that human voice is perceived as more natural and intelligible than synthesised speech \cite{abdulrahman2022natural, obremski2022exploratory}. We are interested in exploring the overall impression of the agent of whether the synthesised speech influences the global perceptive effect of the agent’s human-likeliness.
Further research awaits for study around gesture-generation to create human-like and appropriate ECAs capable of interacting in all types of conversations. The proposed challenge and similar ones can serve as the foundation for the development of such ECAs by identifying and sharing key aspects of gesture generation with the research community.

\begin{acks}
The authors wish to thank Meta Research for the data; Che-Jui Chang for providing the baseline model; Alessandro Sestini, Woo-Ri Ko, Liu Yang for an informal review of this paper. 

This work was partially supported by the Wallenberg AI, Autonomous Systems and Software Program (WASP) funded by the Knut and Alice Wallenberg Foundation;
by the Wallenberg Research Arena (WARA) Media and Language with in-kind contribution from the Electronic Arts (EA) R\&D department, SEED; by Industrial Strategic Technology Development Program (No.\ 20023495) funded by MOTIE, Korea.

\end{acks}

\newpage

\bibliographystyle{ACM-Reference-Format}
\bibliography{main}
\balance

\newpage
\onecolumn 
\appendix
\section{Additional figures}

In this section, we provide additional figures, which could not be included in the main paper due to the space constraint.

\begin{figure}[!t]
\large
    \includegraphics[width=0.9\columnwidth]{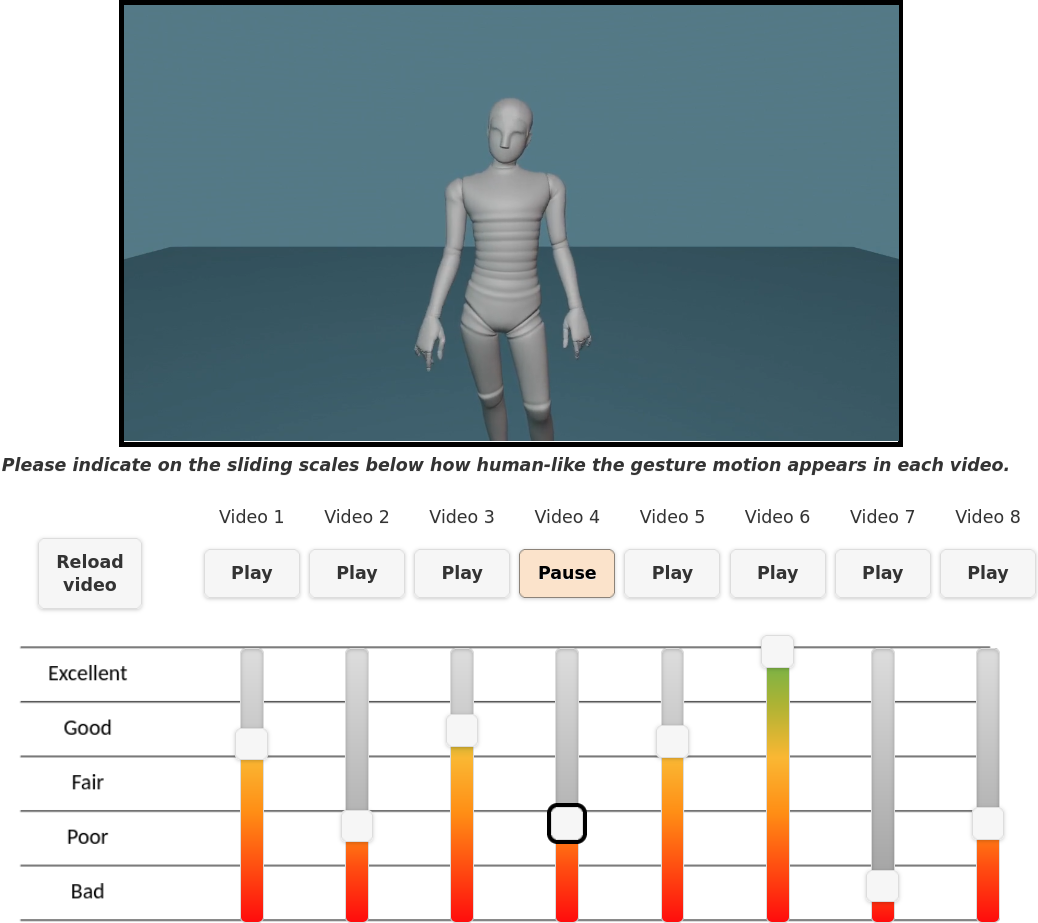}
    \label{sfig:append_ui_page}
\caption{Example screenshot of the modified HEMVIP \cite{jonell2021hemvip} rating interface from the human-likeness evaluation.}
\label{fig:append_hemvipgui}
\end{figure}

\begin{figure*}[!t]
\centering
   \hfill
  \begin{subfigure}[b]{\columnwidth}
    \includegraphics[width=\columnwidth]{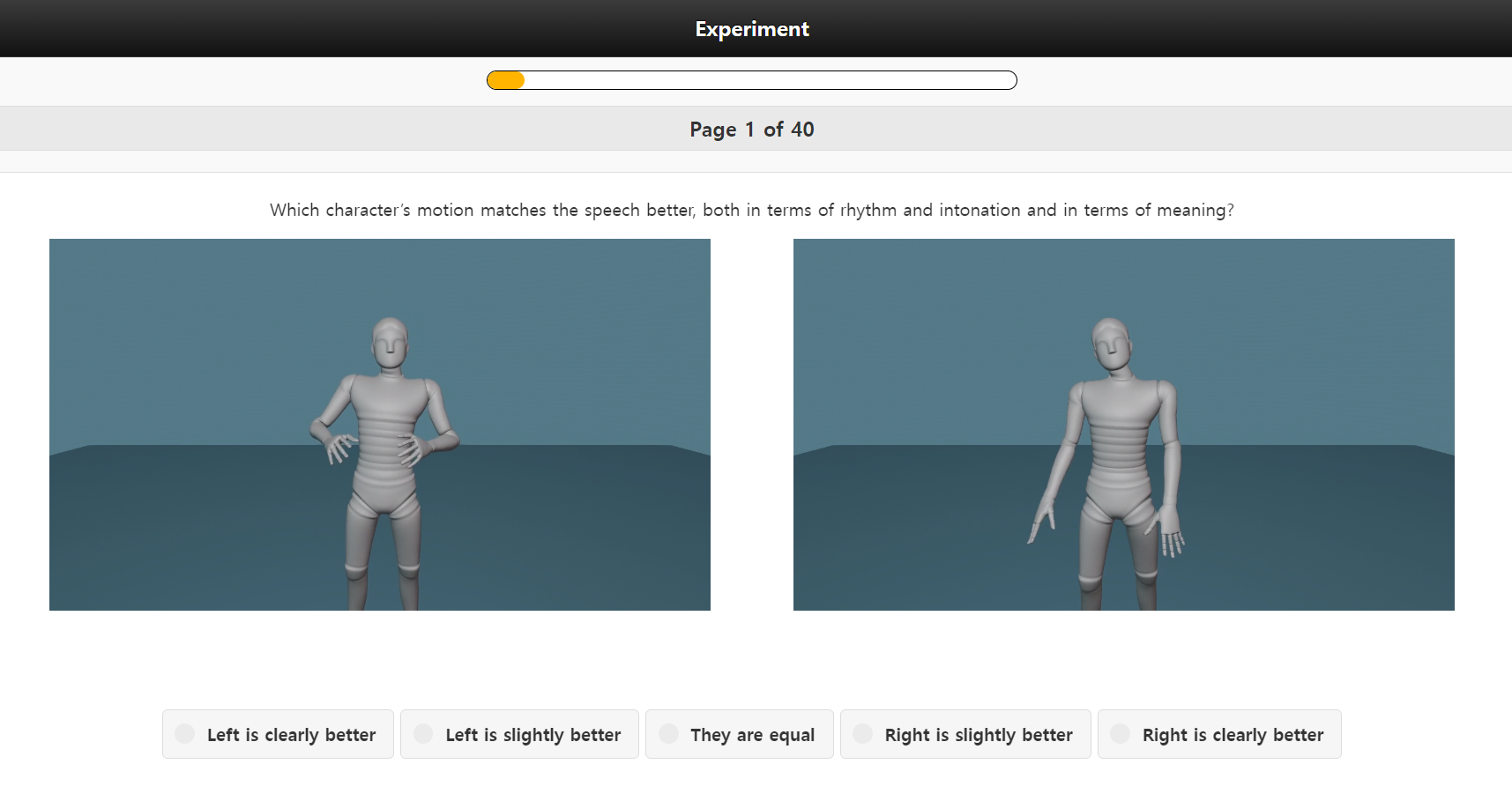}
    \caption{Appropriateness for agent speech}
    \label{sfig:append_monadicgui}
  \end{subfigure}
  \hfill\hfill
  \begin{subfigure}[b]{\columnwidth}
    \centering\includegraphics[width=\columnwidth]{figures/interloc_approp_gui.png}
    \caption{Appropriateness for the interlocutor}
    \label{sfig:append_dyadicgui}
  \end{subfigure}
   \hfill
\caption{Screenshots of the response interfaces from the two appropriateness evaluations, showing questions, response options, and snapshots of stimulus videos for each.}
\label{fig:append_appropgui}
\end{figure*}

Figure \ref{fig:append_hemvipgui} and Figure \ref{fig:append_appropgui} illustrate the GUI (Graphic User Interface) used for the human-likeness and appropriateness studies accordingly.

\section{Additional tables}


Table \ref{tab:append_humlikestats} contains full results of the human-likeness evaluation. Table \ref{tab:append_approp} contains full results of the two appropriateness evaluations.

\begin{table}[!t]
\centering%
\caption{Summary statistics of user-study ratings from the human-likeness study, with confidence intervals at the level $\alpha=0.05$. Conditions are ordered by decreasing sample median rating.}
\label{tab:append_humlikestats}
    \begin{tabular}{@{}lcc@{}}
    \toprule
    Condi- & \multicolumn{2}{c@{}}{Human-likeness} \\ tion & Median & Mean\\
    \midrule
    NA & $71\in{}[70,\,71]$ & $68.4{\pm}1.0$ \\
    SG & $69\in{}[67,\,70]$ & $65.6{\pm}1.4$ \\
    SF & $65\in{}[64,\,67]$ & $63.6{\pm}1.3$ \\
    SJ & $51\in{}[50,\,53]$ & $51.8{\pm}1.3$ \\
    SL & $51\in{}[50,\,51]$ & $50.6{\pm}1.3$ \\
    SE & $50\in{}[49,\,51]$ & $50.9{\pm}1.3$ \\
    SH & $46\in{}[44,\,49]$ & $45.1{\pm}1.5$ \\
    BD & $46\in{}[43,\,47]$ & $45.3{\pm}1.4$ \\
    SD & $45\in{}[43,\,47]$ & $44.7{\pm}1.3$\\
    BM & $43\in{}[42,\,45]$ & $42.9{\pm}1.3$\\
    SI & $40\in{}[39,\,43]$ & $41.4{\pm}1.4$\\
    SK & $37\in{}[35,\,40]$ & $40.2{\pm}1.5$\\
    SA & $30\in{}[29,\,31]$ & $32.0{\pm}1.3$\\
    SB & $24\in{}[23,\,27]$ & $27.4{\pm}1.3$\\
    SC & $\hphantom{0}9\in{}[\hphantom{0}9,\,\hphantom{0}9]$ & $11.6{\pm}0.9$\\
    \bottomrule
    \end{tabular}
\end{table}

\begin{table*}[!t]
\centering%
\caption{Summary statistics of user-study responses from both appropriateness studies, with confidence intervals for the mean appropriateness score (MAS) at the level $\alpha=0.05$. ``Pref.\ matched'' identifies how often test-takers preferred matched motion in terms of appropriateness, ignoring ties. Conditions are ordered by mean appropriateness score.}
\label{tab:append_approp}
\hfill
\begin{subtable}{\columnwidth}
\centering%
\caption{Appropriateness for agent speech}%
\label{stab:append_monadic}%

\begin{tabular}{@{}l|cc|cccccc@{}}
\toprule
Condi- & \multirow{2}{*}{MAS} & Pref. & \multicolumn{6}{c}{Raw response count} \\
tion & & matched & 2 & 1 & 0 & $-$1 & $-$2 & Sum \\\midrule
NA & \hphantom{$-$}0.81$\pm$0.06 & 73.6\% & 755 & 452 & 185 & 217 & 157 & 1766\\
SG & \hphantom{$-$}0.39$\pm$0.07 & 61.8\% & 531 & 486 & 201 & 330 & 259 & 1807\\
SJ & \hphantom{$-$}0.27$\pm$0.06 & 58.4\% & 338 & 521 & 391 & 401 & 155 & 1806\\
BM & \hphantom{$-$}0.20$\pm$0.05 & 56.6\% & 269 & 559 & 390 & 451 & 139 & 1808\\
SF & \hphantom{$-$}0.20$\pm$0.06 & 55.8\% & 397 & 483 & 261 & 421 & 249 & 1811\\
SK & \hphantom{$-$}0.18$\pm$0.06 & 55.6\% & 370 & 491 & 283 & 406 & 252 & 1802\\
SI & \hphantom{$-$}0.16$\pm$0.06 & 55.5\% & 283 & 547 & 342 & 428 & 202 & 1802\\
SE & \hphantom{$-$}0.16$\pm$0.05 & 54.9\% & 221 & 525 & 489 & 453 & 117 & 1805\\
BD & \hphantom{$-$}0.14$\pm$0.06 & 54.8\% & 310 & 505 & 357 & 422 & 220 & 1814\\
SD & \hphantom{$-$}0.14$\pm$0.06 & 55.0\% & 252 & 561 & 350 & 459 & 175 & 1797\\
SB & \hphantom{$-$}0.13$\pm$0.06 & 55.0\% & 320 & 508 & 339 & 386 & 262 & 1815\\
SA & \hphantom{$-$}0.11$\pm$0.06 & 53.6\% & 238 & 495 & 438 & 444 & 162 & 1777\\
SH & \hphantom{$-$}0.09$\pm$0.07 & 52.9\% & 384 & 438 & 258 & 393 & 325 & 1798\\
SL & \hphantom{$-$}0.05$\pm$0.05 & 51.7\% & 200 & 522 & 432 & 491 & 170 & 1815\\
SC & $-$0.02$\pm$0.04 & 49.1\% & \hphantom{0}72 & 284 & 1057 & 314 & \hphantom{0}76 & 1803\\
\bottomrule
\end{tabular}

\end{subtable}%
\hfill\hfill
\begin{subtable}{\columnwidth}
\centering%
\caption{Appropriateness for the interlocutor}%
\label{stab:append_dyadic}%

\begin{tabular}{@{}l|cc|cccccc@{}}
\toprule
Condi- & \multirow{2}{*}{MAS} & Pref. & \multicolumn{6}{c}{Raw response count} \\
tion & & matched & 2 & 1 & 0 & $-$1 & $-$2 & Sum \\
\midrule
NA & \hphantom{$-$}0.63$\pm$0.08 & 67.9\% & 367 & 272 & \hphantom{0}98 & 189 & \hphantom{0}88 & 1014\\
SA & \hphantom{$-$}0.09$\pm$0.06 & 53.5\% & \hphantom{0}77 & 243 & 444 & 194 & \hphantom{0}55 & 1013\\
BD & \hphantom{$-$}0.07$\pm$0.06 & 53.0\% & \hphantom{0}74 & 274 & 374 & 229 & \hphantom{0}59 & 1010\\
SB & \hphantom{$-$}0.07$\pm$0.08 & 51.8\% & 156 & 262 & 206 & 263 & 119 & 1006\\
SL & \hphantom{$-$}0.07$\pm$0.06 & 53.4\% & \hphantom{0}52 & 267 & 439 & 204 & \hphantom{0}47 & 1009\\
SE & \hphantom{$-$}0.05$\pm$0.07 & 51.8\% & \hphantom{0}89 & 305 & 263 & 284 & \hphantom{0}73 & 1014\\
SF & \hphantom{$-$}0.04$\pm$0.06 & 50.9\% & \hphantom{0}94 & 208 & 419 & 208 & \hphantom{0}76 & 1005\\
SI & \hphantom{$-$}0.04$\pm$0.08 & 50.9\% & 147 & 269 & 193 & 269 & 129 & 1007\\
SD & \hphantom{$-$}0.02$\pm$0.07 & 52.2\% & \hphantom{0}85 & 307 & 278 & 241 & 106 & 1017\\
BM & $-$0.01$\pm$0.06 & 49.9\% & \hphantom{0}55 & 212 & 470 & 206 & \hphantom{0}63 & 1006\\
SJ & $-$0.03$\pm$0.05 & 49.1\% & \hphantom{0}31 & 157 & 617 & 168 & \hphantom{0}39 & 1012\\
SC & $-$0.03$\pm$0.05 & 49.1\% & \hphantom{0}34 & 183 & 541 & 190 & \hphantom{0}45 & \hphantom{0}993\\
SK & $-$0.06$\pm$0.09 & 47.4\% & 200 & 227 & 111 & 276 & 205 & 1019\\
SG & $-$0.09$\pm$0.08 & 46.7\% & 140 & 252 & 163 & 293 & 167 & 1015\\
SH & $-$0.21$\pm$0.07 & 44.0\% & \hphantom{0}55 & 237 & 308 & 270 & 144 & 1014\\
\bottomrule
\end{tabular}

\end{subtable}%
\end{table*}

\end{document}